\documentstyle [epsf] {l-aa}

\def\kms{km\thinspace s$^{-1}$ }     
\def\arcs{\ifmmode {'' }\else $'' $\fi}     
\def\arcm{\ifmmode {' }\else $' $\fi}     

\def\etal                {et\thinspace al.}

\def\mv                  {$m_V$}
\def\Mv                  {$M_V$}

\def\pmm                 {$\pm$}

\def\rc                  {$r_c$}

\def\sigccf              {$\sigma_{\rm CCF}$}
\def\sigref              {$\sigma_{\rm ref}$}
\def\sigin               {$\sigma_{\rm in}$}

\def\sigp                {$\sigma_p$}

\def\vr                  {$V_{r}$}

\def\x                   {$\times$}

\def\pxmy#1#2{\setbox0=\hbox{\scriptsize $+#1$}%
\setbox1=\hbox{\scriptsize $-#2$}\dimen0=\ht1%
\advance\dimen0by-1.2pt\,\raise1.3\dimen0%
\copy0\kern-\wd0\lower.7\dimen0\copy1 \,}
\def\xovery#1#2{\setbox0=\hbox{#1}%
\setbox1=\hbox{#2}\dimen0=\ht1%
\advance\dimen0by-1.2pt\,\raise1.2\dimen0%
\copy0\kern-\wd0\lower0.6\dimen0\copy1 \,}
\def\x{$\times$~}
\def\ifigx#1#2{\epsfxsize=#1 \centering{\mbox{\epsfbox{#2}}}}
\def\ifigy#1#2{\epsfysize=#1 \centering{\mbox{\epsfbox{#2}}}}
\def\farcs{\hbox{$.\!\!^{\prime\prime}$}}
\begin{document}

\thesaurus{03(11.09.1 M31; 11.19.4; 11.08.1; 08.11.1; 03.20.7; 03.13.5)}

\title{Velocity dispersions and mass-to-light ratios for 9 globular
clusters in M31}

\author{P.~Dubath\inst{1,2}, Carl~J.~Grillmair\inst{3}}

\institute{Observatoire de Gen\`eve, 
           ch. des Maillettes 51, CH-1290 Sauverny, Switzerland
      \and INTEGRAL Science Data Centre, 
           ch. d'\'Ecogia 16, CH-1290 Versoix, Switzerland 
      \and Jet Propulsion Laboratory, 4800 Oak Grove Drive, 
           Mail Stop 179-225, Pasadena, CA 91109}

\offprints{Pierre.Dubath@obs.unige.ch}

\date{Received , accepted}
\maketitle
\markboth{Dubath \& Grillmair}{Velocity dispersions for M31 globular clusters}


\begin{abstract}

We present internal velocity dispersion determinations from
high-resolution spectroscopic observations of a sample of nine
globular clusters in M31.  Comprehensive numerical simulations are
used to show that the typical uncertainty of our velocity dispersion
measurements is $\sim$5\%. Using these new velocity dispersions
together with structural parameters derived from HST observations, we
estimate the $M/L_V$ ratios of these clusters and find that they are
typical of those measured for Galactic clusters. We show relations
between velocity dispersion, luminosity and physical scales for
globular clusters belonging to the Galaxy, the Magellanic clouds,
Fornax, M31, and Centaurus A. The mean relations and the degree of
scatter are similar in all galaxies. This reveals remarkable
similarities, in term of structure and $M/L_V$ ratio, between the
globular clusters belonging to these different galaxies. We briefly
discuss the possible use of individual globular clusters as
extragalactic distance indicators.

\end{abstract}


\section{Introduction}

Valuable information about globular clusters and galaxy formation can
be obtained by investigating the extent to which the properties of
globular clusters belonging to different galaxies are similar. For
example, differences in initial mass function during cluster formation
and/or in susbsequent cluster dynamical evolution, which may both
depend on galactic environment, would translate into differences in
present-day stellar content. The globular cluster stellar content can
be characterized by the mass-to-light ($M/L$) ratio, which can now be
determined for relatively remote globular clusters.

Collecting all the available data from the literature, Pryor \& Meylan
(1993) derive $M/L_V$ ratios for 56 Galactic globular clusters using
King-Michie dynamical models. They obtain {\it global} $M/L_V$ ratios
ranging from about 1 to 5 with a mean of 2.3 (in solar units), and
found no significant correlations (apart from a possible weak one
between $M/L_V$ and cluster mass) between the {\it global} $M/L_V$
ratios and other parameters such as metallicity, concentration,
half-mass relaxation time, or distance from either the Galactic center
or the Galactic plane. $M/L_V$ ratios similar to those obtained for
Galactic clusters have been obtained in studies of globular clusters
belonging to the Magellanic clouds (Dubath et al.\ 1996b) and to
the Fornax dwarf spheroidal galaxy (Dubath et al.\ 1992).

Another way of investigating globular cluster similarities, in terms
of structure and $M/L_V$ ratio, is to look at the correlations between
velocity dispersion, luminosity and a physical size scale.  These
correlations, which are analogous to the fundamental plane
correlations for elliptical galaxies, have already been discussed for
Galactic clusters by several authors (e.g., Meylan \& Mayor 1986,
Paturel \& Garnier 1992, Djorgovski \& Meylan 1994, Djorgovski 1995).
The tight correlation between the velocity dispersion, the core radius
and the central surface brightness obtained for Galactic clusters
(Djorgovski 1995) is consistent with expectations from the Virial
theorem assuming that Galactic globular cluster cores have a universal
and constant $M/L_V$ ratio to within the measurement errors.

Because of its relative proximity and large size, the M31 globular
cluster system is an obvious target for the study of extragalactic
clusters. Previous studies of various aspects of the M31 globular
cluster system have been reviewed by Fusi Pecci et al.\ (1993), Huchra
(1993), Tripicco (1993), and Cohen (1993). The only
velocity-dispersion determinations of M31 clusters published so far
are by Peterson (1988). Corresponding $M/L_V$ ratio estimates are
given for two clusters and are found to be similar to those typically
obtained for Galactic clusters. A limitation in this work, however,
arises from the difficulty of measuring M31 cluster structural
parameters from the ground. In M31 the angular sizes of core and
half-light radii are typically $\sim$0\farcs2 and $\sim$1\arcs,
respectively.

In this work, we present new velocity dispersion and $M/L_V$ ratio
determinations for a sample of M31 globular clusters, for which
structural parameters derived from HST observations are available in
the literature. The $M/L_V$ ratio estimates are based on simple
relations derived from the Virial theorem and from King models. Our
velocity dispersion estimates are also key observational constraints
for more detailed dynamical analyses, e.g., based on Fokker-Plank or
King-Michie multi-mass models, which are beyond the scope of this
paper.

The spectroscopic observations and the data reduction are presented in
Sect.~2. Numerical simulations used for deriving velocity dispersions
from the integrated-light spectra and the corresponding results are
described in Sect.~3.  Section~4 discusses the structural parameters,
and $M/L_V$ estimates are given in Sect.~5. The relations between
velocity dispersion, luminosity and different physical scales are
discussed in Sect.~6 for our M31 cluster sample together with samples
of clusters belonging to the Galaxy, the Magellanic clouds, the Fornax
dwarf spheroidal galaxy, and Centarus A.  We summarize our findings in
Sect. ~7.

\section{Observations and Data Reductions}

We obtained 12 high-resolution integrated-light spectra of 10 globular
clusters belonging to M31. The observations were made September 9-11,
1994 (see Table~\ref{tabsig}), with the Hamilton echelle spectrograph
(Vogt 1987) operated at the coud\'e focus of the 3-m Shane telescope
of the Lick Observatory. The detector was a thinned,
backside-illuminated TI CCD, with 800 \x\ 800 pixels of 15 \x\ 15
$\mu$m each, and with a readout noise of about 6 electrons.

During the first night, a relatively wide entrance slit of 2\farcs0
\x\ 5\farcs0 was used to match poor seeing conditions. A slit of
1\farcs5 \x\ 5\farcs0 was used during the second and the third
nights. As a consequence, the instrumental resolution, as estimated
from the FWHM of the emission lines of thorium calibration spectra,
was slightly lower during the first night (31\,000 or 9.7 \kms) than
during the second and the third ones (35\,000 or 8.6 \kms).  Both
slits used have widths larger than the apparent half-light radii (see
Table~\ref{tabmol}) of all our clusters.


Table~\ref{tabsig} gives the date of the observations, the exposure
time (Texp) and the signal-to-noise (S/N) ratio of the cluster
spectra. During each observing night, many radial velocity standard
stars were also measured. Spectra of a thorium-argon lamp were taken
6-8 times through each of the three nights.

The spectra were reduced with INTER - TACOS, a new software package
developed in Geneva by D.~Queloz and L.~Weber. The first-night spectra
contain 50 useful orders ranging from 4800 to 8300 \AA, and those
obtain the second and third night contain 49 orders ranging from 4600
to 7500 \AA. The different orders do not overlap, i.e., the echelle
spectra exhibit holes in the wavelength coverage. The orders are never
rebinned, nor merged together.


The shifts in velocity between the different thorium-argon lamp
spectra taken during one particular night are always smaller than 1
\kms. However, a velocity shift as large as 5 \kms\ is observed
between the second and the third nights. As a first step, for each
night, one wavelength solution is computed from a thorium-argon
spectrum, and applied to all the spectra taken during that particular
night. The cluster spectra also contain numerous strong night-sky OH
and O2 emission lines, because of the long exposure time on relatively
faint objects. The residual velocity zero-point shift of each globular
cluster spectrum is then derived by measuring\footnote{The measurement
of the mean velocity shift of the night-sky emission lines is achieved
by cross-correlating the observed spectrum with a template constructed
using the rest wavelengths of the emission lines} the mean shift (in
velocity) of the emission lines compared to their accurate rest
wavelengths (taken from Osterbrock et al.\ 1996). As a second step of
the wavelength calibration, all the {\it globular cluster\/} spectra
are corrected according to these velocity zero-point shifts.


The reduced spectra are cross-correlated with an optimized numerical
mask, used as a template (see Dubath et al.\ 1990 for the details of
the cross-correlation technique). In short, our template contains
lower and upper wavelength limits for each line from a selected sample
of narrow spectral lines. For a particular radial velocity, the value
of the cross-correlation function (CCF) is given by the sum, over all
spectral lines, of the integral of the considered spectrum within the
lower and the upper line limits shifted according to the given
velocity. The CCF is thus built step by step over the velocity range
of interest. The CCF is not affected by the fact that the spectra are
composed of a succession of independent orders, with holes in the
wavelength coverage. The CCF is a kind of mean spectral line over the
approximately 1400 useful lines spread over the spectral ranges of the
different orders. The lower and upper limits of the lines are computed
from a mix of observed and theoretical high-resolution spectra of K2
giants, in such a way as to optimize the CCF (see Dubath et al.\
1996a).

\begin{figure}
\ifigx{8.0cm}{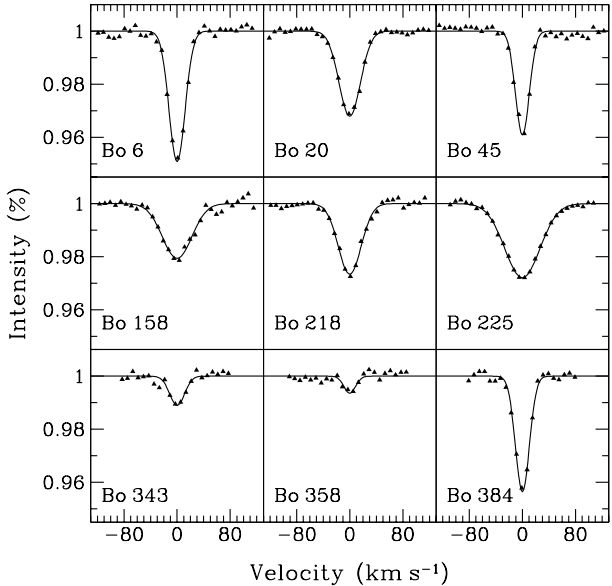}
\caption{Cross-correlation functions (CCFs) of the M31 globular
cluster spectra. The dots represent the CCFs themselves, and the
continuous lines represent the fitted Gaussians. Cluster designations
are from Battistini et al.~(1980).}
\label{ccf}
\end{figure}

When the number of considered lines is large enough ($>$ several
hundreds), this cross-correlation technique produces a CCF which is
nearly a perfect Gaussian. A Gaussian function is fitted to each
derived CCF in order to determine three physical quantities: (1) the
location of the minimum equal to the radial velocity \vr, (2) its
depth $D$, and (3) its standard deviation \sigccf, related to line
broadening mechanisms.

Figure~\ref{ccf} displays the CCFs of the integrated-light spectra
obtained for the M31 globular clusters in our sample.

We have a total of 32 measurements of 11 standard stars, mostly giant
stars of spectral type G8 to K4, collected during the same observing
run. The comparison of the standard-star radial velocities with
reference values provided by CORAVEL measurements (Mayor, private
communication) shows that the instrumental radial velocity accuracy is
of order 0.5 \kms\ and that the zero-point shift between the two
datasets is not significant. In the case of the M31 clusters, the
accuracy of the radial velocity zero-point is very probably even
better owing to our use of the night-sky emission lines to calibrate
the zero point of each cluster spectrum.

The mean value of the widths (\sigccf) of the CCFs obtained for our
sample of standard stars is 5.8\pmm0.3 \kms\ for the first night, and
5.5\pmm0.2 for the two others. This difference is the direct
consequence of the different slit widths used. We see no dependence of
the CCF width on spectral type over the range considered here (G8 to
K4). We also know that the CCF width does not depend on the star
metallicity from previous measurements of a large number of various
types of metal-deficient stars (see Fig.~6 of Dubath et al.\ 1996a).


\section{Velocity dispersions from integrated-light spectra}

The {\it projected\/} velocity dispersions can be derived from the
broadening of the cluster CCFs. This broadening results from the
Doppler line broadening present in the integrated-light spectra
because of the random spatial motions of the stars. Since the
spectrograph slit used is large compared to the the M31 globular
clusters apparent size, a very large number of stars contributes
significantly to the integrated light.  Quantitatively, recent
numerical simulations (Dubath et al.\ 1994, 1996a) show that
statistical errors, which can be very important for integrated-light
measurements of some Galactic globular clusters because of the
dominance of a few bright stars, are negligible in the present
case. An integrated light spectrum of an M31 globular cluster is well
approximated by the convolution of the spectrum of a typical globular
cluster star with the projected velocity distribution.

Since the CCF is a kind of mean spectral line, the above property also
hold for CCFs. The CCF of an integrated-light spectrum of a globular
cluster is the convolution of the CCF of the spectrum of a typical
globular cluster star by the projected velocity distribution. Since
the CCFs have a Gaussian shape, an estimate of the projected velocity
dispersion \sigp\ in the integration area of a globular cluster is
given by the quadratic difference,

\begin{equation}
       \sigma_p^{\,2} = 
       \sigma_{\rm CCF}^{\,2}{\rm\, (cluster)} - 
       \sigma_{\rm ref}^{\,2},
\label{sigest}
\end{equation}

where \sigccf\ is the width of the Gaussian fitted to the cluster CCF
and \sigref\ is the average width of the \sigccf\ obtained for a
sample of standard stars, as representative as possible of the cluster
stars which contribute most to the integrated light.

We do not use this formula directly in the present study. In order to
derive the cluster projected velocity dispersion, we carry out a large
number of numerical simulations. The results of these simulations
confirm, however, the validity of Eq.~\ref{sigest} (see
Sect.~\ref{gene}).

\subsection{Numerical Simulations \label{numsim}}

\begin{figure*}
\ifigy{12.3cm}{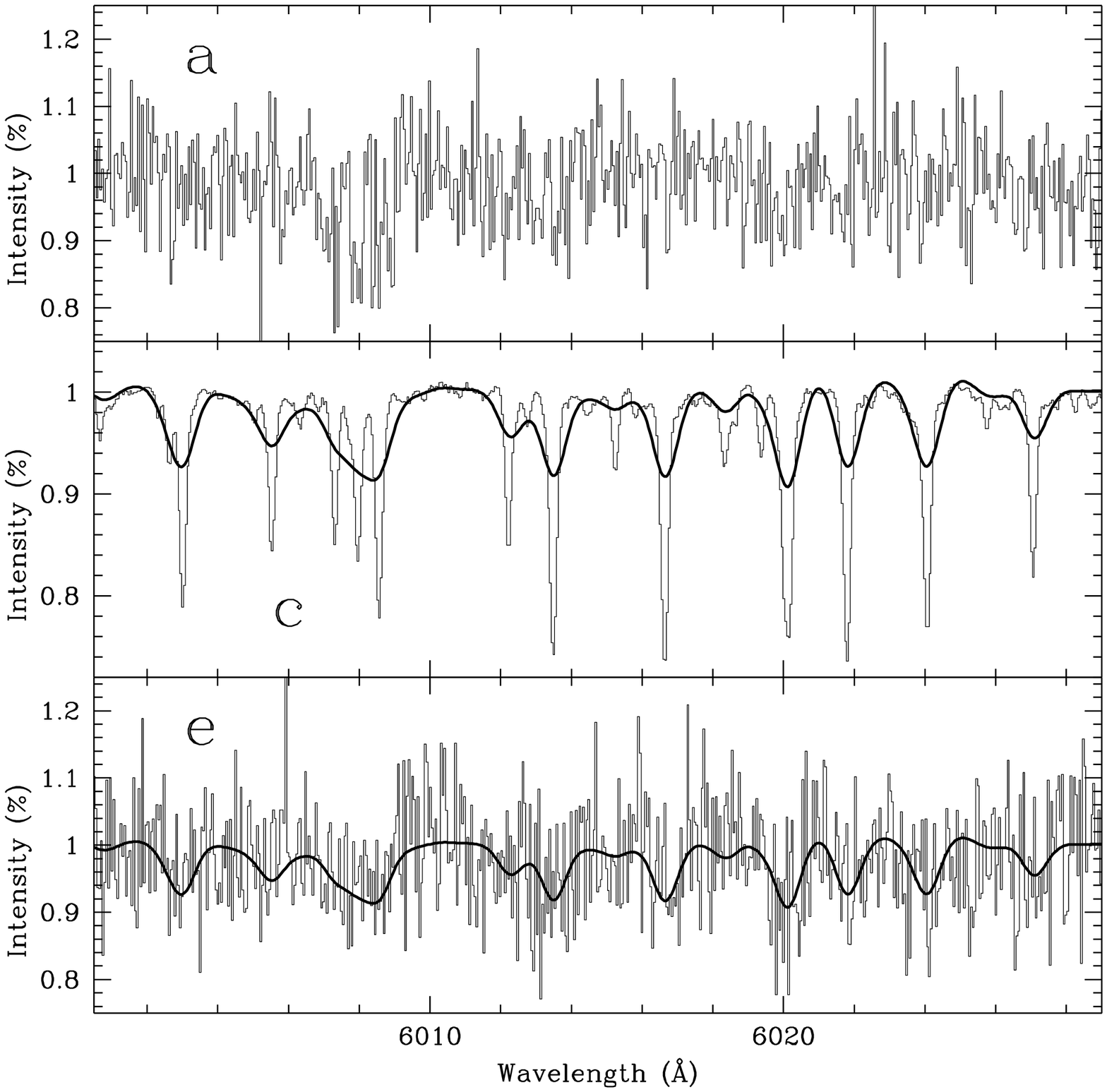}\hspace*{0.6cm}\ifigy{12.3cm}{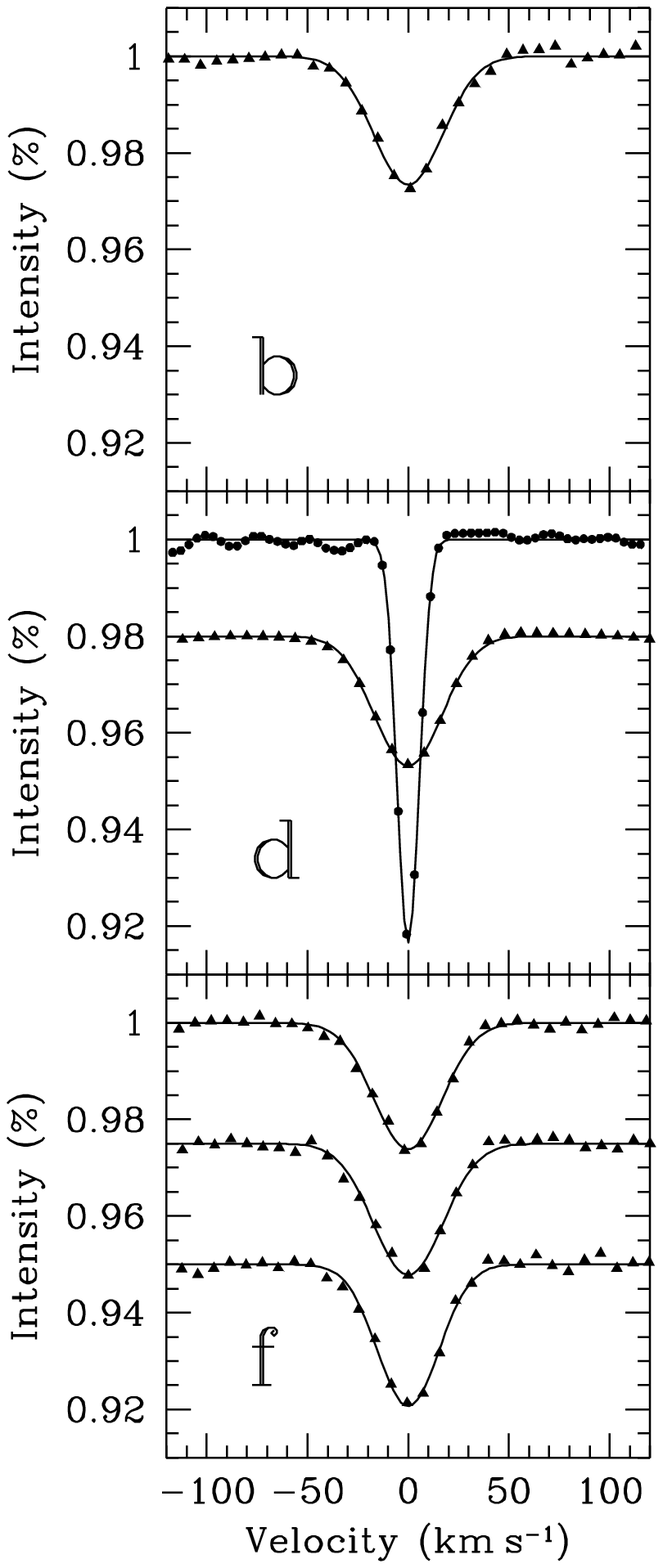}
\caption{Illustration of the numerical simulations in the case of the
cluster Bo~218. This figure displays: (a) the {\it observed\/}
integrated-light spectrum of the cluster; (b) its cross-correlation
function (CCF); (c) the original and the convolved standard-star
template spectrum; (d) their corresponding CCFs, one arbitrarily
shifted for the purpose of display; (e) the convolved template
spectrum (again) and one example of this spectrum with additional
noise, simulating the {\it observed\/} cluster spectrum (panel a); and
(f) three examples of CCFs of noisy convolved spectra taken at random
and arbitrarily shifted, simulating the cluster CCF (panel b). This
figure shows only a tiny fraction of the spectra ($\sim$1/50 of their
total wavelength range).}
\label{simuill}
\end{figure*}

In order to simulate the integrated-light spectrum and the CCF
obtained for globular cluster, we proceed in several steps. In the
first row of Fig.~\ref{simuill}, the integrated-light spectrum (left)
and the CCF (right) obtained for the cluster Bo~218 are
displayed. Corresponding simulation results are displayed in the last
row, while the intermediate steps of the simulation are illustrated in
the middle row of this figure.

1. The simulation input parameter is the cluster velocity dispersion
(\sigin). The simulation starts with a high signal-to-noise spectrum
of a standard star of appropriate spectral type, e.g., a K2 giant. To
simulate the Doppler line broadening, this spectrum is convolved with
a Gaussian function of standard deviation equal to the input velocity
dispersion \sigin. A portion of the standard-star spectrum, before and
after convolution, is shown in the panel (c) of Fig.~\ref{simuill}.

2. In general, the spectral lines of the convolved standard-star
spectrum do not have the same depths as those of a given cluster
spectrum, mainly because of possible metallicity difference. The next
step is thus to adjust the depth of the standard-star spectral lines.
This is done by scaling the convolved standard-star spectrum so that
its CCF is of the same depth as the cluster CCF.  (With our
cross-correlation technique, there is a clear relationship between the
average depth of the spectral lines of a spectrum and the depth of the
spectrum CCF). In Fig.~\ref{simuill} for example, the spectra and the
CCFs displayed in the middle row are linearly scaled so that the
convolved spectrum CCF (the broad CCF in panel d) be of the same depth
as the cluster CCF (in panel b). A noiseless template of a cluster
spectrum -- for one particular velocity dispersion \sigin\ -- results
from the second step.

3. Random noise is added to this template spectrum to simulate the
photon counting and CCD readout noises of observed cluster spectra.  A
portion of the template spectrum for Bo~218 (c) and of one example of
simulated noisy spectrum are displayed in panel (e) of
Fig.~\ref{simuill}. We then cross-correlate the simulated noisy
spectrum and derive the radial velocity (\vr) and the sigma (\sigccf)
of the resulting CCF. This third step is repeated a large number of
times (100 to 150 times) to observe the influence of the noise on \vr\
and \sigccf. Panel (f) of Fig.~\ref{simuill} shows three examples of
simulated CCFs taken at random. The comparison of the upper and the
lower panels of this figure shows how well our simulations reproduce
the integrated-light spectrum and the CCF obtained for Bo~218.

\begin{figure}
\ifigx{7.0cm}{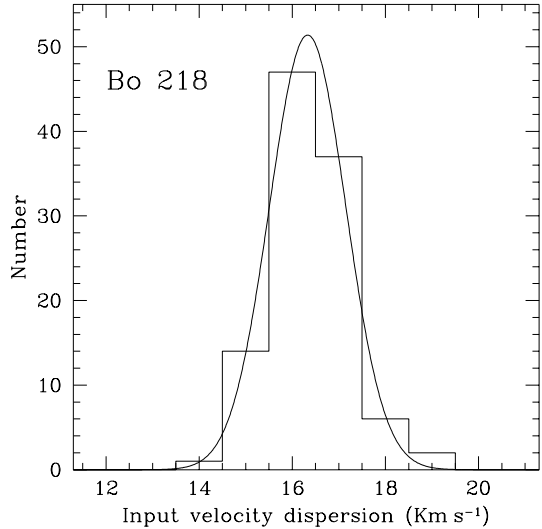}
\caption{Distribution of the number of time that the simulations
produce a CCF consistent -- in terms of width -- with the observed CCF
as a function of the input velocity dispersion (\sigin), in the case
of Bo~218. For each \sigin, 100 simulations are considered. The
continuous line represents a Gaussian fitted to the distribution.}
\label{simudis}
\end{figure}

In order to derive the best estimates of the projected velocity
dispersion (\sigp) in the clusters, we proceed as follows.  For each
cluster, a set of different input velocity dispersions (\sigin),
varying by step of 0.5 or 1 \kms\ around the expected cluster ``true''
velocity dispersion (first evaluated with Eq.~\ref{sigest}) are
considered. For each \sigin, 100 or 150 simulations are carried out
and the relative number of times that the width of the simulated CCFs
is consistent with the width of the observed CCF is reported.
Consistent means here that the sigma of the simulated CCF must be
within \pmm\ 0.2-0.5 \kms\ (depending on the different cases) of the
sigma of the observed CCF.  Figure~\ref{simudis} shows, in the case of
Bo~218, the distribution of these numbers as a function of the input
velocity dispersions (\sigin) of the simulations. This distribution is
a kind of probability distribution; the best estimate of the ``true''
cluster velocity dispersion is given by the \sigin\ which leads most
often, through the simulations, to a CCF consistent with the observed
one. For example, among 100 simulations carried out with \sigin\ = 16
\kms, 47 have a width consistent with the width of the observed CCF
obtained for Bo~218, while none of the 100 simulations with \sigin\ =
13 \kms\ is successful in reproducing the observed CCF.

A similar distribution is derived for each cluster, and a Gaussian is
fitted to each of them. The resulting means and sigmas, which provide
the most probable cluster velocity dispersions (\sigp) and their
uncertainties, are given in column (11) of Table~\ref{tabsig}.

\subsubsection{Generalization \label{gene}}

This section presents a generalization of the simulation results
(which can be skipped by less concerned readers). For each of the 9
clusters, simulations are carried out for a set of input velocity
dispersions (\sigin). In the simulations, a particular cluster is
characterized by the spectrum S/N ratio and by the depth ($D$) of the
cluster CCF. Therefore in general, the input parameters are the
spectrum S/N, the CCF depth $D$, and the input velocity dispersion
\sigin. For each set of input parameters, 100 to 150 simulated CCFs
are computed, and distributions of the resulting radial velocities and
CCF sigmas are thus produced. The standard deviations of these
distributions -- $\varepsilon$(\vr) for the radial velocities and
$\varepsilon$(\sigccf) for the sigmas -- provide estimates of the
uncertainties due to the spectrum noise. The following formula,

\begin{equation}
       \varepsilon\, = \, \frac{C}{{\rm S/N}\, D} \, \left( 1+\alpha
\sigma_{\rm in} \right)  
\label{epsi}
\end{equation}

where $C$ and $\alpha$ are two constants, is fitted to the results of
the simulations. With $C = 0.09$ and $\alpha = 0.06$, this equation
gives in \kms, (1) $\varepsilon$(\vr) with an accuracy of order of
10\%, and (2) $\varepsilon$(\sigccf)) with an accuracy of about
20\%. It can then be used to estimated the uncertainties due to the
spectrum noise in a more general way, for any set of parameters S/N,
$D$, and \sigin.  Equation~\ref{epsi} is a generalization of Eq.~(3)
of Dubath et al. (1990) to broad CCFs. The constant $C$ is lower in
the present paper than in this previous study because the larger
wavelength range, and consequently larger number of spectral lines,
taken into account in the cross-correlation process ($C$ scales
roughly with the square root of the number of lines).

\begin{figure}[ht]
\ifigx{7.0cm}{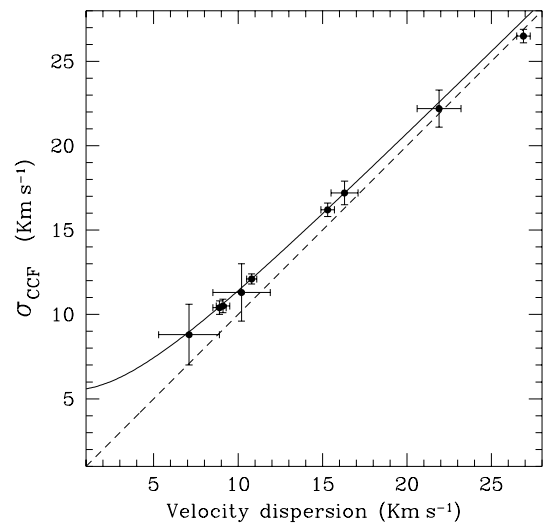}
\caption{Relation between the sigmas (\sigccf) of the observed CCFs
and the best estimates of the velocity dispersions (\sigp) resulting
from the simulations. Each point displays the result for one of the 9
M31 clusters. The continuous line represents
Eq.~(\protect\ref{sigest}), whose validity is thus confirmed by the
results of the simulations.}
\label{simusig}
\end{figure}

Figure~\ref{simusig} displays, for each cluster, the sigma (\sigccf)
of the CCFs of the observed spectrum as a function of the
corresponding estimates of the velocity dispersion (\sigp) resulting
from the simulations. Eq.~(\ref{sigest}) is illustrated in this figure
by the continuous line, and the good agreement with the simulation
results indicates that this equation is a valid model.

\subsection{Results}

\begin{table*}
\caption[]{Radial velocities and velocity dispersions of M31 globular clusters}
\begin{flushleft}
\begin{tabular}{rrcccrrrcrr}
\noalign{\smallskip}
\hline
\noalign{\smallskip}
\multicolumn{2}{l}{Cluster}& \mv\ & [Fe/H] & Date      & \multicolumn{1}{c}{Texp}  & S/N & \multicolumn{1}{c}{\vr}   & $D$ &\multicolumn{1}{c}{\sigccf}&\multicolumn{1}{c}{\sigp} \\
\multicolumn{2}{r}{number} &      &        & 94~Sept.  & \multicolumn{1}{c}{(min)} &     & \multicolumn{1}{c}{(\kms)}& (\%)&\multicolumn{1}{c}{(\kms)} &\multicolumn{1}{c}{(\kms)} \\
(1)&(2)                    & (3)  &  (4)   & (5)       & \multicolumn{1}{c}{(6)}   & (7) & \multicolumn{1}{c}{(8)}   & (9) &\multicolumn{1}{c}{(10)}    &\multicolumn{1}{c}{(11)} \\
\noalign{\smallskip}
\hline
\noalign{\smallskip}
  6    & 58    & 15.8  & $-$0.57 & ~9 & 100~~ & 13 & $-$236.9\pmm0.2 & 4.9 & 12.1\pmm0.3  & 10.6\pmm0.4 \\ 
 20    & 73    & 14.6  & $-$1.07 & 11 & 100~~ & 18 & $-$350.8\pmm0.3 & 3.2 & 16.2\pmm0.4  & 15.3\pmm0.5 \\ 
 45    & 108   & 15.8  & $-$0.94 & ~9 & ~85~~ & 11 & $-$425.1\pmm0.3 & 3.9 & 10.4\pmm0.4  &  8.7\pmm0.5 \\ 
147    & 199   & 15.5  & $-$0.24 & 11 & ~80~~ & 26 &  $-$50.7\pmm0.1 & 7.3 &  6.0\pmm0.3  &    star?~~~~\\
158    & 213   & 14.5  & $-$1.08 & 10 & ~50~~ & 10 & $-$186.5\pmm1.0 & 2.1 & 22.2\pmm1.1  & 21.9\pmm1.3 \\
218    & 272   & 14.7  & $-$1.18 & 11 & ~50~~ & 11 & $-$220.2\pmm0.6 & 2.7 & 17.2\pmm0.7  & 16.3\pmm0.8 \\ 
225    & 280   & 14.3  & $-$0.70 & ~9 & 120~~ & 31 & $-$164.8\pmm0.3 & 2.8 & 26.5\pmm0.4  & 26.9\pmm0.5 \\ 
343    & 105   & 16.3  & $-$1.49 & 11 & 100~~ &  8 & $-$359.6\pmm1.7 & 1.1 & 11.3\pmm1.7  & 10.2\pmm1.7 \\ 
{\tt "}&{\tt "}&{\tt "}&{\tt "}  & 10 & ~63~~ &  4 & $-$357.5\pmm4.0 & 0.7 &  7.7\pmm4.1  & ...~~~~~ \\ 
358    & 219   & 15.1  & $-$1.83 & 10 & ~90~~ & 12 & $-$315.1\pmm1.8 & 0.6 &  8.8\pmm1.8  &  7.1\pmm1.8 \\ 
{\tt "}&{\tt "}&{\tt "}&{\tt "}  & ~9 & ~50~~ &  6 & $-$317.5\pmm3.1 & 0.8 & 12.0\pmm3.1  & ...~~~~~ \\ 
384    & 319   & 15.7  & $-$0.66 & 11 & 100~~ & 10 & $-$363.8\pmm0.3 & 4.4 & 10.5\pmm0.4  &  9.1\pmm0.5 \\ 
\noalign{\smallskip}
\hline
\end{tabular} 
\label{tabsig}
\end{flushleft}
\end{table*}

For each observation, Table~\ref{tabsig} lists the cluster
identification from Battistini et al.\ (1980) in column~(1), and from
Sargent et al.\ (1977) in column~(2), the cluster apparent V magnitude
in column~(3), the cluster [Fe/H] in column~(4), the date of the
observation in column~(5), the exposure time in column~(6), the
signal-to-noise ratio of the integrated-light spectrum in column~(7),
the heliocentric radial velocity in column~(8), the depth of the CCF
in column~(9), the sigma of the CCF in column~(10), and the projected
velocity dispersion in column~(11). The cluster apparent V magnitudes
are taken from Battistini et al.\ (1987), and [Fe/H] are taken from
Huchra et al.\ (1991).

The radial velocity errors given in column (8) of Table~\ref{tabsig}
are computed using Eq.~(\ref{epsi}). This equation provides an
estimate of the error due to the spectrum noise (photon counting and
CCD readout noises) but does not take into account the uncertainty of
the zero-point corrections computed from measurements of the night-sky
emission lines.  Consequently, the radial velocity errors smaller than
$\sim$0.3 \kms\ are probably underestimated. The \sigccf\ errors given
in column (8) of Table~\ref{tabsig} are the square root of the
quadratic sum of the errors due to the noise (estimated using
Eq.~(\ref{epsi})), and of an instrumental error of 0.25 \kms, derived
from standard-star measurements.

Two cluster spectra are too noisy to provide useful velocity
dispersion estimate. The CCF obtained for Bo~147 is as narrow as the
CCFs obtained for individual standard stars, and much narrower that
the value expected from its absolute magnitude.  Consequently, Bo~147
is almost certainly a foreground star.



\section{Structural parameters} 

\begin{table*}
\caption[]{Mass-Luminosity Ratios for M31 Globular Clusters}
\begin{flushleft}
\begin{tabular}{rrccccccccc}
\noalign{\smallskip}
\hline
\noalign{\smallskip}
\multicolumn{2}{l}{Cluster}& $r_c$ & $r_{hp}$ & $c$ & M$_V$ & $M_{cl}$ & $M_{cl}^{Virial}$ & E(B-V) & $M/L_V$ & $M/L_{V}^{Virial}$ \\
\multicolumn{2}{c}{ID} & pc & pc &  & & M$_{\odot}$ & M$_{\odot}$ &  & M$_{\odot}/$L$_{\odot, V}$ & M$_{\odot}/$L$_{\odot, V}$ \\
(1)&(2) & (3) & (4) & (5) & (6) & (7) & (8) & (9) & (10) & (11) \\
\noalign{\smallskip}
\hline
\noalign{\smallskip}
  6 & 58  & 0.62 & 2.8 & 1.76 & -9.0  & $4.6 \times 10^5$ & $7.3 \times 10^5$ & 0.11 & 1.4 & 2.2 \\
 20 & 73  & 1.18 & 4.7 & 1.71 & -10.2 & $1.7 \times 10^6$ & $2.6 \times 10^6$ & 0.11 & 1.7 & 2.5 \\
 45 & 108 & 0.92 & 3.6 & 1.70 & -9.0  & $4.2 \times 10^5$ & $6.3 \times 10^5$ & 0.12 & 1.2 & 1.8 \\
158 & 213 & 1.54 & 5.0 & 1.59 & -10.2 & $3.8 \times 10^6$ & $5.6 \times 10^6$ & 0.08 & 3.7 & 5.5 \\
218 & 272 & 0.47 & 2.0 & 1.74 & -10.0 & $7.9 \times 10^5$ & $1.2 \times 10^6$ & 0.09 & 0.9 & 1.4 \\
225 & 280 & 1.42 & 4.9 & 1.63 & -10.4 & $5.5 \times 10^6$ & $8.2 \times 10^6$ & 0.09 & 4.4 & 6.5 \\
343 & 105 & 0.34 & 2.9 & 2.02 & -8.3  & $3.7 \times 10^5$ & $7.0 \times 10^5$ & 0.06 & 2.1 & 4.0 \\
358 & 219 & 1.82 & 4.7 & 1.46 & -9.5  & $4.0 \times 10^5$ & $5.5 \times 10^5$ & 0.06 & 0.7 & 1.0 \\
384 & 319 & 0.57 & 2.2 & 1.69 & -8.9  & $2.8 \times 10^5$ & $4.3 \times 10^5$ & 0.06 & 0.9 & 1.4 \\
\noalign{\smallskip}
\hline
\end{tabular} 
\label{tabmol}
\end{flushleft}
\end{table*}

M31 is at a distance where the core radii of typical globular
clusters are considerably smaller in angular size than the typical,
ground-based seeing disk. This makes measurements of $r_c$ difficult,
as attested to by the large range of values obtained by different
authors for the same clusters. Fortunately, six of the clusters in our
sample have recently been observed with the Hubble Space Telescope
(HST), and accurate values of $r_c$ have been published (Fusi-Pecci
\etal~ 1994 [Bo218, 384]; Grillmair \etal~ 1996 [Bo6, 45, 343, 358]).
The values obtained span a large range but are typical for the
brighter globulars in our own Galaxy. For the remaining 3 clusters in
Table 1, we use the ground-based measurements of Battistini \etal~
(1982) and Crampton \etal~ (1985). For two clusters (Bo6, 45) common
to all three samples, the agreement between the ground-based values
and those measured by Grillmair \etal is typically quite good, with
the worst cases differing by no more than 30\%.

The surface density profiles of these clusters have also been
difficult to characterize from the ground owing the very low surface
brightness relative to the M31 background and the inability to resolve
individual stars. However with the advent of HST we now know that
globular clusters in M31 are structurally similar to those found in
our own Galaxy. These clusters can generally be characterized by King
models, though departures from King models in the form of collapsed
cores (Bo343, Bendinelli \etal~ 1993; Grillmair \etal 1996) and tidal
tails (Bo6, 343, 358, Grillmair \etal~ 1996) have also been found. In
all respects, it seems that globular clusters in M31 as a group are of
the same breed as the clusters belonging to our own Galaxy.

For the clusters observed by Grillmair \etal (1996), we have
determined the half-light radii $r_{hp}$ (the radii within which half
the total light of the clusters is contained in projection) by
integrating over King models with appropriate core radii and
concentration parameters ($c = \log r_t/r_c$). For the 2 clusters 
observed by Fusi Pecci \etal~ (1994) we have used their
core and half-light radii and integrated over a grid of King models to
infer the corresponding concentration parameter. For the three clusters
for which we have only ground-based measurements of $r_c$, we follow
Battistini \etal~ (1982) in assuming a uniform value of $r_t = 60$pc.
Our adopted values of $r_c$ and $r_{hp}$ are tabulated in Table 2. In
all cases, the published values of $r_c$ have been scaled in
accordance with our adopted distance to M31 of 770kpc (Ajhar \etal~
1996).


\section{Mass-to-light ratios}

If we assume that the clusters are reasonably well represented
by King models then, following Queloz \etal~ (1995), we can compute
the total mass using

\begin{equation}
	M_{cl} = {9 \over 2 \pi G}{ \mu r_c \sigma_p^2(0) \over \alpha p}.
\label{mass}
\end{equation}

\noindent Values of $\mu, \alpha$, and $p$ have been tabulated for a
range of concentration parameters by King (1966) and Peterson \&
King (1975). 

The $\sigma_p(0)$ in Eq.~\ref{mass} refers to the {\it central}
velocity dispersion of the cluster.  Given the small angular size of
the clusters and our use of 1.5 and 2.0 arcsecond-wide slits, it is
clear that our dispersion measurements will have been influenced by
light from moderately large radii. We tested the effect of our
slit-widths on the measured velocity dispersions by integrating
luminosity-weighted, King model velocity dispersion profiles over the
area subtended by our slit. We found that, over a large range in $c$,
and for both slit-widths used, the measured velocity dispersion would
be lower than the central velocity dispersion by about 5\%. Hence, for
the purposes of computing masses, we increased the measured
dispersions in Table \ref{tabsig} accordingly. The resulting cluster
masses we obtain using Eq.~\ref{mass} are listed in column 7 of Table
\ref{tabmol}.

An alternative to the somewhat model-specific method used
above is a straightforward application of the Virial theorem:

\begin{equation}
	M_{cl} \simeq {4 \sigma_p^2 r_{hp} \over 0.4 G}, 
\label{virial}
\end{equation}

\noindent where we have assumed an isotropic velocity distribution and 
$r_{hp} \approx {4 \over 3} r_h$ (Spitzer 1987), where $r_h$ is the
half-mass radius. The masses computed using this equation are given in
column 8 of Table \ref{tabmol}.

Owing to M31's low Galactic latitude, obscuration of the globular
clusters by foreground Galactic dust varies significantly from one
side of M31 to the other. Using the extinction maps of Burstein \&
Heiles (1982), E(B-V) was estimated for each cluster and is given in
column 9 of Table \ref{tabmol}. Using the $V$-magnitudes given by
Battistini \etal~ (1987), we adopt $A_V$ = 3.2 E(B-V) (DaCosta \&
Armandroff 1990), and $(m-M_V) = 24.432$ to compute total cluster
$V$-band luminosities. The corresponding values for $M/L_V$ are given
in column 10 and 11 of Table \ref{tabmol}.

The $M/L_V$ ratios given in Table \ref{tabmol} are remarkably similar
to those typically found in Galactic globulars (Pryor \& Meylan 1993).
Bo158 and 225 might seem a trifle high, but we note that both these
clusters have only ground-based measurements of $r_c$. It is entirely
possible that these estimates of $r_c$ suffer from incomplete removal
of the effects of seeing and are consequently too high. The largest
source of uncertainty in $M/L_V$ is generally in the estimation of
$r_c$, being of the order of 15\% even for the HST-imaged
clusters. Uncertainties in the magnitude estimates of Battistini
\etal, in our estimates of the local extinction, and in the velocity
dispersion measurements contribute $\leq 10\%$ each to the final
uncertainty. Bo343 and 358 are exceptions to this general rule, having
reasonably well-measured core radii, but rather less well-determined
velocity dispersions. The formally estimated uncertainties in $M/L_V$
are $\approx 20\%$ for those clusters observed with HST, and probably
closer to 50\% for those clusters imaged only from the ground.

Table~\ref{tabmol} shows that the $M/L_V$ ratios derived with
Eq.~\ref{virial} (column 11) are systematically $\sim$50\% larger than
those obtained with Eq.~\ref{mass} (column 10). This gives a rough
idea on how model dependent are our estimates.

\section{Relation between velocity dispersion, luminosity, and a 
physical scale}

\begin{figure*} 
\ifigx{16cm}{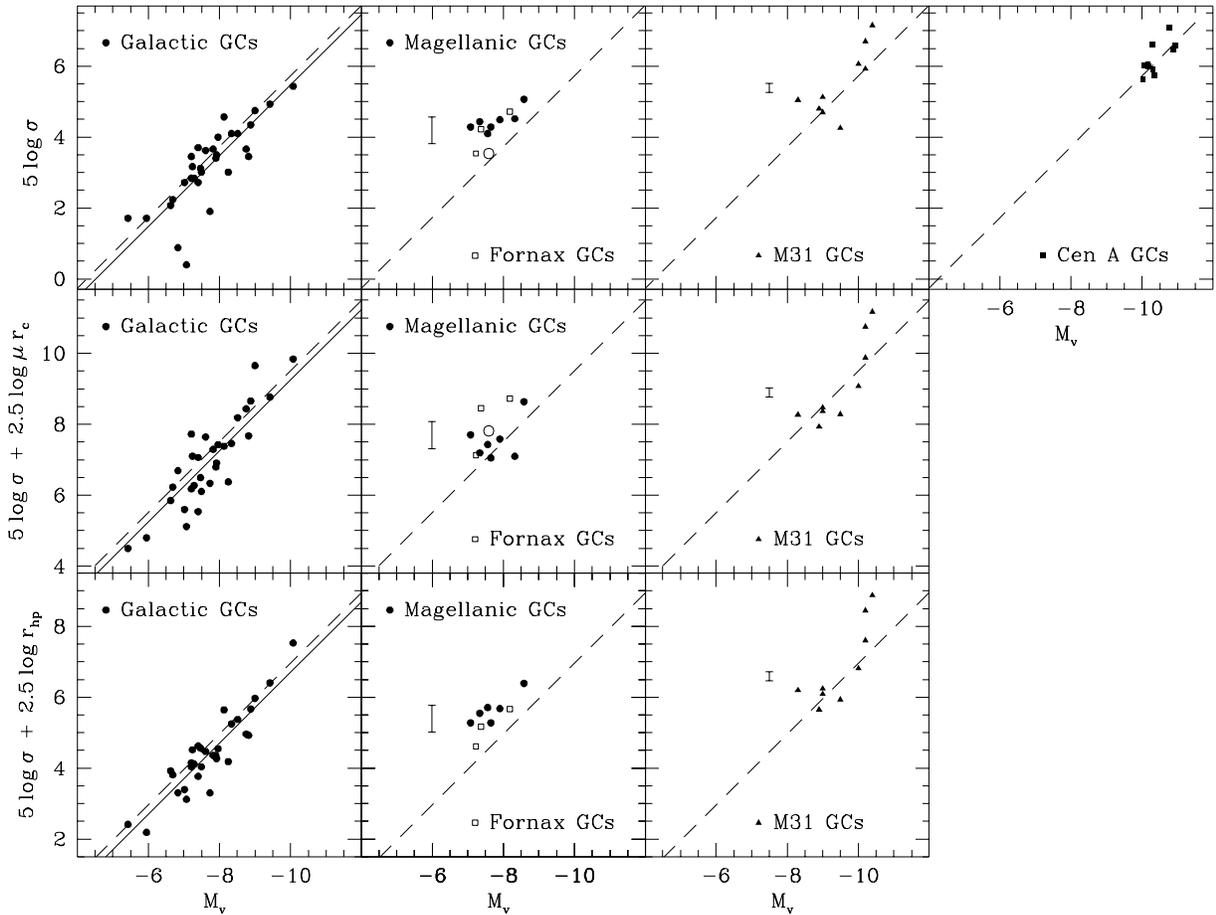}
\caption{Different projections of the fundamental plane for Galactic,
old Magellanic, Fornax, M31, and Centaurus A globular clusters.  Shown
are the velocity dispersion \sigp\ (uppermost row of panels), and
combinations of \sigp\ with physical scales -- the product of the
dimensionless mass $\mu$ and the core radius \rc\ (second row) and
half-light radius $r_{hp}$ (third row) -- as a function of the
absolute visual magnitude \Mv. The straight lines are derived by
fitting the relations derived from King models (Eq.~\ref{King_2}) or
the Virial Theorem (Eq.~\ref{Virial_2}) to the {\it Galactic} clusters
data (left panels) only, using the {\it global} (continuous lines) or
the {\it central} (dashed lines) velocity dispersions.  Consequently,
the different dashed lines in each row are identical and are displayed
for comparison with the data. The circle represents NGC~121, the only
SMC cluster. The error-bars represent 7\pmm1.2 \kms\ in the case of
Magellanic clusters and 12\pmm0.7 \kms\ in the case of M31 clusters,
and illustrate the typical errors of our \sigp\ measurements}
\label{fp}
\end{figure*}

Another way to investigate similarities among different globular
cluster populations is to look at the relation between velocity
dispersion, luminosity, and a physical size scale. In such a parameter
space, systematic differences between two systems of globular
clusters, in terms of structure or $M/L$ ratio, would result in either
different mean relations or in different degrees of scatter about a
given relation. We do not attempt here to derive mean relations
through bivariate fits (e.g. Djorgovski 1995) as they turn out to be
rather unstable because of both the small number of data points as
well as the weak correlation between the physical size parameter and
the luminosity and velocity dispersion. Instead, we fit to the data
the relations expected from the Virial theorem, or from the King
models.  Assuming a constant $M/L_V$ ratio, the Virial Theorem
predicts the relation

\begin{equation}
5 \log \sigma\, +\, 2.5 \log r_{hp}\, \sim \, -\,M_V
\label{Virial_2}
\end{equation}

\noindent
between the {\it global} velocity dispersion $\sigma$, the half-light
radius $r_{hp}$, and the absolute visual magnitude \Mv. Similarly, 
King models predict

\begin{equation}
5 \log \sigma_0\, +\, 2.5 \log \mu\, r_c\, \sim \, -\,M_V,
\label{King_2}
\end{equation}

\noindent where $\sigma_0$ is the {\it central} velocity dispersion,
$\mu$ a dimensionless parameter which varies with cluster
concentration $c$, $r_c$ the core radius, and \Mv\ the absolute visual
magnitude.  Figure~\ref{fp} shows the relations between $5 \log
\sigma$ vs. $M_V$ (uppermost row of panels), $5 \log \sigma\, +\, 2.5
\log r_h$ vs.  $M_V$ (second row), and $5 \log \sigma\, +\, 2.5 \log
\mu\, r_c$ vs.  $M_V$ (third row), for different data sets. For the
Galactic clusters we use velocity dispersion measurements based on
radial velocities of individual stars, taken from the compilation of
Pryor and Meylan (1993). For the other clusters, we use velocity
dispersions derived from integrated-light observations, taken from
Dubath et al.\ (1996a) for 8 old Magellanic clusters, from Dubath et
al.\ (1992) for 3 clusters belonging to the Fornax dwarf spheroidal
galaxy, from the present study for 9 M31 clusters, and from Dubath
(1994) for 10 Centaurus A clusters. The continuous lines
represent the relation $5\log\sigma\,=\,Cst\,-\,M_V$ (uppermost row),
and the relations (\ref{Virial_2}) and (\ref{King_2}), in the second
and third rows respectively, with constants derived by fitting the
{\it Galactic} cluster data. The dashed lines show the relations
obtained when central velocity dispersions (extrapolated for these
clusters by Pryor and Meylan [1993] using King models) are considered
instead of the {\it global} velocity dispersion.

It is worth mentioning that correlations resulting from bivariate fits
do not differ much from the relations expected from the Virial theorem
(Djorgovski 1995), and that Fig.~\ref{fp} would look very similar if
slightly different projections were used. The scatter of the data
points probably results to a large extent from measurement errors, and
it is quite remarkable that this scatter is of the same order in all
panels of Fig.~\ref{fp}. The similarity of the different panels of
Fig.~\ref{fp} indicates both that there is no large systematic 
differences in globular cluster $M/L_V$ ratios between one galaxy to
another, and that measurement errors are comparable, and even smaller,
for extragalactic clusters than for Galactic clusters.

In the first row of Fig.~\ref{fp}, we expect a higher degree of
scatter since the physical scale of each cluster (which acts as a
second parameter) is not taken into account. Notably, a few
large-size/low-concentration Galactic clusters lie well below the
other clusters in the upper left panel. The fact that similar large
clusters are not present in our M31 and Cen A samples is due to a
selection bias which favors brighter and generally more compact
clusters.  In any case, the relatively small spread of the data points
around the straight lines in the uppermost panels points to an
additional similarity in terms of physical size range between the
clusters of these different galaxies, with the possible exception of
the old Magellanic clusters. This is particularly interesting in the
case of the Centaurus A clusters since these clusters are brighter
than the Galactic ones, but their physical scales have not yet been
measured.

For the old Magellanic clusters, the products $\mu$ $r_c$ are
systematically {\it smaller} than the Galactic average. This can
perfectly explain why the Magellanic clusters appear above the
Galactic relation in the upper panel in Fig.~\ref{fp}, while the
Magellanic and Galactic middle panels are similar. The lower panel is,
however, rather puzzling. Unexpectedly, the data points all lie above
the Galactic relation, as if the half-light radii used here and taken
from van den Bergh (1994) were about 40\% too large. This point is
further discussed in another paper (Dubath et al.\ 1996b).

The two M31 clusters (Bo158 \& 225) with large $M/L_V$ ratio estimates
from last section stand out above the Galactic relation in
Fig.~\ref{fp}. As already pointed out, these clusters only have
ground-based measurements of $r_c$ which may be overestimates.  The
scatter of the other clusters is remarkably small compared to the
scatter of Galactic clusters.

\subsection{Individual globular clusters as extragalactic distance 
indicators?}

As illustrated in Fig.~\ref{fp}, the absolute magnitude of an
individual globular cluster can be derived from its velocity
dispersion and physical scale with an accuracy of $\sim$0.5 magnitude.
For a given parent galaxy, providing there is no systematic
differences in $M/L_V$, an accurate distance modulus can in principle
be computed using the mean distance modulus of a moderate number (10
-- 20) of its associated globular clusters. In other word, plotting
apparent instead of absolute magnitude in Fig.~\ref{fp}, the
difference between the distance modulii of two parent galaxies is
given by the horizontal shift required to bring the two (dereddened)
data sets in agreement.  Combining the capabilities of the HST and of
10-m class, ground-based telescopes, this method could be applied for
galaxies out as far as the Virgo cluster.

\section{Summary}

In this paper, we present projected velocity dispersion measurements
from integrated-light spectra for a sample of 9 M31 globular clusters.
By means of comprehensive numerical simulations, we show that the
typical relative uncertainty of our measurements is $\sim$5\%.
Because of the relatively large distance of M31 and the aperture
angular size used our integrated-light observations sample a large
fraction of these very bright clusters. Consequently, statistical
uncertainties due to small samples of bright dominant stars which can
affect integrated-light measurements of nearby Galactic clusters are
completely negligible. This is confirmed by numerical simulations
presented in Dubath et al.\ (1996a).  Paradoxically, it is in some
respects easier to measure the {\it global} velocity dispersion of a
M31 cluster than that of a Galactic clusters. For the brightest M31
clusters, reliable velocity dispersions can be measured with a
3-m-class telescope using single exposures of order one hour long.

Previous velocity dispersion measurements are available from Peterson
(1988) for 3 of our clusters. The agreement is good for the cluster
Bo20, while we obtain significantly lower and more accurate values for
Bo158 and 225.

Combining the new velocity dispersions with structural parameters, we
compute the cluster $M/L_V$ ratios using relations derived from King
models and the Virial theorem. These $M/L_V$ ratios appear remarkably
similar to those found for Galactic clusters (see e.g., Pryor \&
Meylan 1993).  Two clusters (Bo158 \& 225) having only ground-based
measurements of $r_c$ have $M/L_V$ ratio estimates somewhat above the
range of Galactic values.  It is entirely possible that these
estimates of $r_c$ suffer from incomplete removal of the effects of
seeing and are consequently too high.

Another way to investigate similarities of globular clusters located
in different parent galaxies is to look at the relation between
velocity dispersion, luminosity and a physical scale
(Fig~\ref{fp}). Using additional data from previous papers, we find
remarkable similarities, in terms of $M/L_V$ ratios and structures,
between the globular clusters located in our Galaxy, the Magellanic
clouds, the Fornax dwarf spheroidal, M31, and Centaurus A.  It is
worth emphasizing that our samples include some of the brightest M31
and Centaurus A clusters, which are both brighter and more massive
than any Galactic cluster, as expected because of the larger number of
clusters in both these galaxies.

Our M31 cluster sample is not large enough to investigate possible
correlations between the $M/L_V$ ratio estimates and other cluster
parameters. Only a weak correlation is observed with Galactic data
(Pryor \& Meylan 1993), and it is possible that the variations of the
current $M/L_V$ estimates from one cluster to another result to a
large extent from measurement errors.

\begin{acknowledgements} P.D.\ acknowledge support through a grant from 
the Swiss National Science Foundation. P.D.\ warmly thanks Graeme
Smith for providing free use of his workstation during a postdoctoral
stay at the Lick Observatory.

\end{acknowledgements}

{\it Note added in proof.--} As this paper was beeing accepted for
publication, we became aware of a similar work by Djorgovski et
al. (ApJL, in press). They derive velocity dispersions from Keck
observations for a sample of 21 M31 globular clusters, which includes
our 9 clusters. Their results are in good agreement with ours, e.g.,
the difference in velocity dispersion values is always $\leq$ 1.4
\kms\ for all clusters. They point out an interesting trend between
the ratio mass-to-luminosity in the K band and the cluster
metallicity, which has not been observed before.


\begin{thebibliography}{}

\bibitem{a} Ajhar, E.A, Grillmair, C.J., Lauer, T.R., Baum, W.A.,
           Faber, S.M., Holtzman, J.A., Lynds, C.R., \& O'Neil, E.J. Jr. 1996, AJ, 111, 1110
\bibitem{b} Battistini, P., B\`onoli, F., Braccesi, A., Federici, L., Fusi Pecci, F.,
           Marano, B., \& Borngen, F. 1987, A\&AS, 67, 447
\bibitem{c} Battistini, P., B\`onoli, F., Braccesi, A., Fusi Pecci, F.,
           Malagnini, M. \& Marano, B. 1980, A\&AS, 42, 357
\bibitem{d} Battistini, P., B\`onoli, F., Buonanno, R.,
           Corsi, C.E., \& Fusi Pecci, F., 1982, A\&A, 113, 39
\bibitem{e} Bendinelli, O., Cacciari, C., Djorgovski, S., Federici, L,
	   Ferraro, F.R., Fusi Pecci, F., Parmeggiani, G., Weir, N., \& Zavatti,
	   F. 1993, ApJ Letters, 409, L17
\bibitem{f} Burstein, D., \& Heiles, C. 1982, AJ, 87, 1165
\bibitem{g} Cohen, J.G. 1993
           in ASP Conf. Ser., Vol. 48, The Globular Cluster -- Galaxy Connection, 
           G. Smith \& J. Brodie(San Francisco: ASP), 438
\bibitem{h} Crampton, D., Schade, D.J., Chayer, P., \& Cowley, A.P. 1985, ApJ, 288, 494
\bibitem{i} DaCosta, G.S., \& Armandroff, T.E. 1990, AJ, 100, 162
\bibitem{j} Djorgovski, S., \& Meylan, G. 1994, AJ, 108, 1292
\bibitem{k} Djorgovski, S. 1995, ApJL, 438, L29 
\bibitem{l} Dubath, P. 1994, BAAS, 185, 5203
\bibitem{m} Dubath, P., Meylan, G., Mayor, M., \& Magain, P. 1990, A\&A, 239, 142                 
\bibitem{n} Dubath, P., Meylan, G., \& Mayor, M. 1992, ApJ, 400, 510 
\bibitem{o} Dubath, P., Meylan, G., \& Mayor, M. 1994, ApJ, 426, 192 
\bibitem{p} Dubath, P., Meylan, G., \& Mayor, M. 1996a, A\&A, in press 
\bibitem{q} Dubath, P., Meylan, G., \& Mayor, M. 1996b, A\&A, submitted 
\bibitem{r} Fusi Pecci, F., Cacciari, C., Federici, L., \& Pasquali, A. 1993, 
           in ASP Conf. Ser., Vol. 48, The Globular Cluster -- Galaxy Connection, 
           G. Smith \& J. Brodie(San Francisco: ASP), 410
\bibitem{s} Fusi Pecci, F., Battistini, P., Bendinelli,
           O., Bonoli, F., Cacciari, C., Djorgovski, S., Federici, L., Ferraro,
           F.R., Parmeggiani, G., Weir, N., \& Zavatti, F. 1994, A\&A, 284, 349
\bibitem{t} Grillmair, C.J., Ajhar, E.A., Faber, S.M., Baum, W.A.,
           Holtzman, J.A., Lauer, T.R., Lynds, C.R., \& O`Neil, E.J. Jr. 1996, AJ, 111, 2293
\bibitem{u} Huchra, J. 1993, 
           in ASP Conf. Ser., Vol. 48, The Globular Cluster -- Galaxy Connection, 
           G. Smith \& J. Brodie(San Francisco: ASP), 420
\bibitem{v} Huchra, J., Brodie, J.P., \& Kent, S.M. 1991, ApJ, 370, 495
\bibitem{w} King, I.R. 1966, AJ, 71,64
\bibitem{x} Meylan, G., \& Mayor, M. 1986, A\&A, 166, 122
\bibitem{y} Osterbrock, D.E., Fulbright, J.P., Martel, A.R., Keane, M.J., Trager, S.C., 
           \& Basri, G. 1996, PASP, 108, 277
\bibitem{z} Paturel, G., \& Garnier, R. 1992, A\&A, 254, 93
\bibitem{aa} Peterson, C.J., \& King, I.R. 1975, AJ, 80, 427
\bibitem{ab} Peterson, R. 1988, in Dynamique of Dense Stellar System, ed. D. Merritt(Cambridge: 
           Cambridge University Press), 161
\bibitem{ac} Pryor, C., \& Melan, G. 1993, in ASP Conf. Ser., Vol. 50, Structure and
           Dynamics of Globular Clusters, S. Djorgovski \& G. Melan(San Francisco: ASP), 357
\bibitem{ad} Queloz, D., Dubath, P., \& Pasquini, L. 1995, A\&A, 300, 310
\bibitem{ae} Sargent, W.L., Kowal, C.T., Hartwick, F.D.A., \& van den Bergh, S. 1977, AJ, 82, 947
\bibitem{af} Spitzer, L. 1987, {\it Dynamical Evolution of Globular Clusters}, Princeton University 
           Press, Princeton.
\bibitem{ag} Tripicco, M. 1993,
           in ASP Conf. Ser., Vol. 48, The Globular Cluster -- Galaxy Connection, 
           G. Smith \& J. Brodie(San Francisco: ASP), 432
\bibitem{ah} van den Bergh, S. 1994, AJ, 108, 2145 
\bibitem{ai} Vogt, S. 1987, PASP, 99, 1214

\end{thebibliography}
\end{document}